\documentclass[aps,prd,preprintnumbers,nofootinbib,floatfix,floats,groupedaddress,twocolumn]{revtex4}

\usepackage{bm}
\usepackage[utf8]{inputenc}
\usepackage{latexsym}
\usepackage{dcolumn}
\usepackage{amsmath,amsfonts,amssymb}
\usepackage{graphicx,epsfig}
\usepackage{color}
\usepackage[active]{srcltx}
\usepackage[caption=false]{subfig}
\usepackage{slashed}
\usepackage{hyperref}
\usepackage{tikz}
\usetikzlibrary{shapes,snakes}
\usepackage{multirow}
\usepackage{appendix}
\usepackage[mathscr]{euscript}
\usepackage{mathrsfs}
\def\nn{\nonumber}

\def\l{\left}
\def\r{\right}
\def\DM{\mathrm{d}}

\newcommand{\gae}{\lower 3pt \hbox{$\,\, \buildrel {\scriptstyle >}\over {\scriptstyle
\sim}\,\,$}}
\newcommand{\lae}{\lower 2pt \hbox{$\, \buildrel {\scriptstyle <}\over {\scriptstyle
\sim}\,$}}

\reversemarginpar

\begin{document}

\title{Covariant formulation of Generalised Uncertainty Principle}

 \author{Raghvendra Singh}
 \email{raghvendra@imsc.res.in}
 \affiliation{Institute of Mathematical Sciences, Homi Bhabha National Institute (HBNI), IV Cross Road, ~C. I. T. Campus, Taramani, Chennai 600 113, India}

 \author{Dawood Kothawala}
 \email{dawood@iitm.ac.in}
 \affiliation{Centre for Strings, Gravitation and Cosmology, Department of Physics, Indian Institute of Technology Madras, Chennai 600 036}

\date{\today}
\begin{abstract}
\noindent
We present a formulation of the generalised uncertainty principle based on commutator $\l[ {\hat x}^i, {\hat p}_j \r]$ between position and momentum operators defined in a covariant manner using normal coordinates. We show how any such commutator can acquire corrections if the momentum space is curved. The correction is completely determined by the extrinsic curvature of the surface $p^2=$ constant in the momentum space, and results in non-commutativity of normal position coordinates $\l[ {\hat x}^i, {\hat x}^j \r] \neq 0$. We then provide a construction for the momentum space geometry as a suitable four dimensional extension of a geometry conformal to the three dimensional relativistic velocity space - the Lobachevsky space - whose curvature is determined by the dispersion relation $F(p^2)=-m^2$, with $F(x)=x$ yielding the standard Heisenberg algebra. 
\end{abstract}

\maketitle
\vskip 0.5 in
\noindent
\maketitle
\section{Introduction} \label{sec:intro} 
The Heisenberg algebra satisfied by position and momentum operators, ${\hat x}^i$ and ${\hat p}_j$, forms the cornerstone of quantum mechanics, and in standard treatments, this is characterized by the commutators:
\begin{eqnarray}
\l[ {\hat x}^i, {\hat p}_j \r] &=& i \hbar \delta^i_{\phantom{i}j}
\nn , \\
\l[ {\hat p}_i, {\hat p}_j \r] &=& 0
\nn , \\
\l[ {\hat x}^i, {\hat x}^j \r] &=& 0.
\end{eqnarray}
However, the generalisation and interpretation of the above commutators in a fully relativistic theory which is also invariant under general coordinate transformations is not straightforward. The key issue is already evident from the explicit appearance of ${\hat x}^i$ in the commutator relationships, which spoils manifest general covariance. This issue needs to be addressed since general covariance is also the first step towards generalization of a theory to a curved background space, or spacetime. Therefore, a curved space(-time) generalisation of the Heisenberg algebra requires a covariant definition of the quantities appearing in the above relations. This, in itself, is not difficult if one identifies the coordinates appearing in the commutators via exponential map - the so called normal coordinates. These coordinates, by definition, depend on the geodesic spray from a given point ${\mathcal P}_0$, and hence their characterization depends directly on the curvature of the manifold. We will discuss this in Sec. \ref{sec:commutator}.

A second generalization of the Heisenberg algebra, that has gained considerable interest over the past two decades \cite{Kempf} is tied to what we expect from measurements at very high energies or at very small length scales. Very general arguments rooted in basic principles of quantum mechanics and general relativity suggest that Planck scale might provide a fundamental operational bound on the measurement of observables such as spacetime intervals \cite{paddy}. One therefore expects the above commutators to be modified at large momentum, with Planck scale as the characteristic scale at which the modification occurs. This has motivated a large amount of research exploring a modification of the form 
\begin{eqnarray}
\l[ {\hat x}^i, {\hat p}_j \r] &=& i \hbar \Theta^i_{\phantom{i}j} ({\hat p}_k),
\end{eqnarray}
with the most popular form of the function $\Theta$ being a quadratic in ${\hat p}_k$ that reduces to $\delta^i_{\phantom{i}j}$ at low momentum \cite{GUP}. However, although there exists several arguments based on certain well motivated requirements that have been used to constrain $\Theta^i_{\phantom{i}j}$ at least in flat spacetime (using, for instance, translation and rotation invariance, see \cite{Kempf2}), no such generic constraints are available for a curved spacetime. 

Our first aim in this paper will be to provide a geometrically well motivated argument that gives a curved spacetime generalization of the Heisenberg algebra, in terms of quantities defined covariantly in a specific region of spacetime, provided only that it is a geodesically convex neighbourhood of some spacetime point. Our analysis will also give us a hint towards a geometrical interpretation of $\Theta^i_{\phantom{i}j} ({\hat p}_k)$, constructing which will be our next aim.

As a first step towards explicitly constructing $\Theta^i_{\phantom{i}j} ({\hat p}_k)$, we will begin by first analysing the geometry of the momentum space. To do this, we will start with the well-known expression for center-of-mass energy of a system of two point particles with rest masses $m_1$ and $m_2$, and use it to obtain a metric on the three dimensional momentum space which turns out to be conformal to the well-known Lobachevsky metric on the relativistic velocity space. We then extend this metric to four dimensions by demanding that, for particles with zero relative velocities, it gives the (squared) difference of rest masses: $(\Delta m)^2 = (m_2-m_1)^2$. The momentum space so constructed is flat (${\rm \bf Riemann} =0$) as long as the relation $p^2=g^{ab} p_a p_b = - m^2$ holds for a given $\left(p_a, m\right)$. This also yields $\Theta^i_{\phantom{i}j} ({\hat p}_k)=\delta^i_{\phantom{i}j}$. For a general dispersion relation $F(p^2)=-m^2$, ${\rm \bf Riemann} \neq 0$ and we show that $\Theta^i_{\phantom{i}j} ({\hat p}_k)$ is now determined by the extrinsic geometry of $p^2=$ constant surface in momentum space.

To summarise, we present in this paper a novel formulation of the generalised uncertainty principle that incorporates and interconnects three key ideas: (i) covariant characterization of ${\hat x}^i$ appearing in the uncertainty principle, (ii) momentum space geometry based on the relativistic velocity space, and (iii) modified dispersion relation $F(p^2)=-m^2$. For some earlier work that comments on similar issues, see \cite{Sabine}. Note, however, that our formalism involves only Lorentz invariant modifications of the dispersion relation, incorporated via a very specific construction of the four dimensional momentum space geometry. Our geometrical approach is also somewhat closer in spirit to Ref. \cite{Wagner}. However, the conceptual set-up and mathematical implementation here are completely different. 

\textit{Conventions:} Throughout the paper, Latin indices (serif/non-serif) run from 0 to 3, and the sans-serif font represent frame indices. Boldface quantities represent four vectors.
\section{Covariant characterizations of commutators} \label{sec:commutator} 

\subsection{Normal coordinates}

\begin{figure}[!h]%
    {{\includegraphics[width=0.5\textwidth]{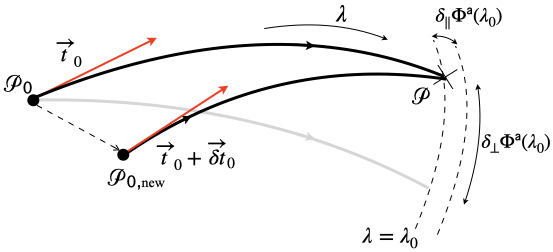} }}%
    \caption{Normal coordinates defined in the tangent space ${ T}_{{\cal P}_0}(M)$. Their variations are completely determined by the geometry of the equi-geodesic surfaces 
    anchored at $\mathcal{P}_0$ (dashed lines), with $\lambda$ the length of the geodesic connecting $\mathcal{P}$ to $\mathcal{P}_0$. 
    The passive version of these variations correspond to shifting the base point $\mathcal{P}_0$, in which case the change in normal coordinates is determined by 
    extrinsic geometry anchored at ${\mathcal P}$.}%
    \label{fig:normal-coords}%
\end{figure}
By definition, the normal coordinates are defined by
$
	\Phi^{\textsf a} = - \eta^{\textsf{ab}} e^{i}_{\textsf b} \nabla_{i} \Omega
$
where $e^{i}_{\textsf b}$ is an orthonormal tetrad at ${\cal P}_0$: $\bm e_{\textsf a} \cdot \bm e_{\textsf b} = \eta_{\textsf{ab}}$ (lorentzian) or $\delta_{\textsf{ab}}$ (euclidean), and $\Omega({\cal P}_0, {\cal P}) = \sigma^2({\cal P}_0, {\cal P})/2$ is the Synge world function \cite{book-jls} with $\sigma^2({\cal P}_0, {\cal P}) = \pm \lambda^2$; see Fig. \ref{fig:normal-coords}. In terms of frame components of the tangent vector, this is equivalent to assigning to ${\cal P}$ the coordinates
$
\Phi^{\textsf a}(\mathcal{P}) = \lambda t^{\textsf a} \left(\mathcal{P}_0\right)
$. The variations in these coordinates are therefore given by \cite{Dewitt Book}
\begin{eqnarray}
\delta \Phi^{\textsf a}(\mathcal{P}) = 
t^{\textsf a} \left(\mathcal{P}_0\right) \l[ \delta \lambda \r]_{t^{\textsf a} \; {\mathrm{fixed}}  }
+ 
\lambda \l[ \delta t^{\textsf a} \left(\mathcal{P}_0\right) \r]_{\lambda \; {\mathrm{fixed}}  }.
\end{eqnarray}
We now wish to compute the variation which arises due to a shift in the origin ${\cal P}_0$, which we characterise by a vector $\varepsilon^{\sf k}$ in ${ T}_{{\cal P}_0}(M)$. The above expression can then be written in a neat form by observing that the variation appearing in the second term on the RHS lies purely within the equi-geodesic surface $\lambda=$ constant, and is therefore given by
\begin{eqnarray}
\l[ \delta t^{\textsf a} \left(\mathcal{P}_0\right) \r]_{\lambda \; {\mathrm{fixed}}  } = K^{\textsf a}_{\phantom{\textsf a} \textsf b} \varepsilon^{\textsf b},
\end{eqnarray}
\\
where $K^{\textsf a}_{\phantom{\textsf a} \textsf b}$ represents the frame components of the extrinsic curvature tensor associated with the equi-geodesic surfaces, and we have used $K^{\textsf a}_{\phantom{\textsf a} \textsf b} t^{\textsf b}=0$. We can therefore write
\begin{eqnarray}
\delta \Phi^{\textsf a}(\mathcal{P}) = \left( \lambda K^{\textsf a}_{\phantom{\textsf a} \textsf b} - t^{\textsf a} t_{\textsf b} \right) \varepsilon^{\textsf b}.
\label{eq:variation-generator}
\end{eqnarray}
The extrinsinc curvature $K^{\textsf a}_{\phantom{\textsf a} \textsf b}$ of the equi-geodesic surfaces is completely determined by curvature tensor of $M$, and it's expansion in normal coordinates can be found in Appendix A of the second reference in \cite{Dawood}.

\subsection{The commutator}

The discussion so far has been general and applies to any manifold. For instance, it can be applied to the case when $M$ represents curved spacetime, and moving to quantum mechanics by identifying the variation of $\Phi^{\textsf a} \equiv {\sf x}^{\textsf a}$ with the commutator of ${\sf x}^{\textsf a}$ with the generator of this variation, which can then be identified with the momentum operator \cite{Kempf3}. Alternatively, when $M$ represents the momentum space,  the variation of $\Phi_{\textsf a} \equiv {\sf p}_{\textsf a}$ can again be identified, quantum mechanically, with the commutator ${\sf p}_{\textsf a}$ with the generator of this variation, which we can then identify with the position operator. 

When $M$ represents the spacetime manifold, the normal coordinates $\Phi^{\textsf a}(\mathcal{P}) \equiv {\sf x}^{\textsf a}$, and the RHS of Eq. (\ref{eq:variation-generator}) suggests the following for the commutator between 
${\sf x}^{\textsf a}$ and the operator $\widetilde{\sf p}_{\textsf b}$ generating the shift of origin by $\varepsilon^{\textsf b}$:
\begin{eqnarray}
\l[ {\sf x}^{\textsf a}, \widetilde{\sf p}_{\textsf b} \r] \overset{\rm def}{:=} i \hbar
\left( \lambda K^{\textsf a}_{\phantom{\textsf a} \textsf b} - u^{\textsf a} u_{\textsf b} \right).
\label{eq:commutator-position}
\end{eqnarray}
where $t^{\textsf a}=u^{\textsf a}$ is the four velocity. In flat spacetime, it follows from the geometry of the equi-geodesic surfaces, that $\lambda K^{\textsf a}_{\phantom{\textsf a} \textsf b} - u^{\textsf a} u_{\textsf b} = \delta^{\textsf a}_{\phantom{\textsf a} \textsf b}$ \cite{poisson, Dawood} and one recovers the standard Heisenberg commutators. On curved spacetimes, the above arguments would imply that the Heisenberg commutators are modified to $[\textsf x^{\textsf a},  \widetilde{\textsf p}_{\textsf b}]=i\hbar \Theta^{\textsf a}_{\ \textsf b}(\boldsymbol{\textsf x})$. However, demanding that the candidate for momentum operator be symmetric adds certain constraints, which we briefly discuss; a very detailed analysis can be found in the seminal papers by De Witt \cite{Dewitt}. To begin with, it is easy to show that the symmetric momentum operator in position representation with respect to the integration measure i.e. $\DM^n x\sqrt{-g}$, is given by 
\begin{eqnarray}
\widetilde{\textsf p}^{\rm{sym}}_{ \textsf a} = -i\hbar(\frac{1}{2}\Theta^{\textsf m}_{\ \textsf a}\Gamma^{\textsf i}_{\ \textsf m \textsf i}+\frac{1}{2}\Theta^{\textsf m}_{\ \textsf a, \textsf m}+\Theta^{\textsf m}_{\ \textsf a}\partial_{\textsf m}).
\end{eqnarray}
The standard commutator relations
\begin{eqnarray}
\l[ {\textsf x}^{\textsf a}, {\textsf p}_{\textsf b} \r] := i \hbar \delta^{\textsf a}_{\phantom{\textsf a} \textsf b},
\label{eq:commutator-std}
\end{eqnarray}
can be obtained even in curved spacetime if we define $\textsf p_{\textsf a}=(\Theta^{-1})^{\textsf b}_{\ \textsf a} \widetilde{\sf p_b}$. But the new momentum operator is no more symmetric with respect to the integration measure. As soon as one tries to make it symmetric, it takes the following form
\begin{eqnarray}
{\sf p}^{\rm{sym}}_{\sf a} =-i\hbar \l(\partial_{\sf a} + \frac{1}{2}\Gamma^{\sf i}_{\phantom{\sf i} \sf a  \sf i} \r) .
\end{eqnarray}

The point we want to emphasize is that, one does not need to modify the commutator relations even in curved spacetime if `$\Theta$' is a function of `$\textsf{x}$' only.
As stated above, this is consistent with known results about quantisation in curved spaces; see, for instance, \cite{Dewitt}. 

Since one expects quantum gravitational effects to play some role in modifying the conventional Heisenberg algebra and hence the uncertainty principle, it is reasonable to expect the deformations to occur at large momenta (suitably identified). We will therefore now turn to the case when $M$ represents the momentum space, and look for the kind of deformations that the geometry of the momentum space can produce in the commutators. To do so, we will now identify the the normal coordinates $(- {\sf x}^{\sf a})$ as generating shifts in ${\sf p}_{\sf a}$ \cite{generators-flows}. To compute the computers, we need information about the equi-geodesic surface and its extrinsic curvature in the momentum space, and for this, we need to first characterise the geometry of the momentum space. We will do so in such a manner that the standard dispersion relation $p^2=-m^2$ will represent. 
\section{Geometry of momentum space} \label{sec:momentum-space}

\begin{figure*}
    {{\includegraphics[width=0.90\textwidth]{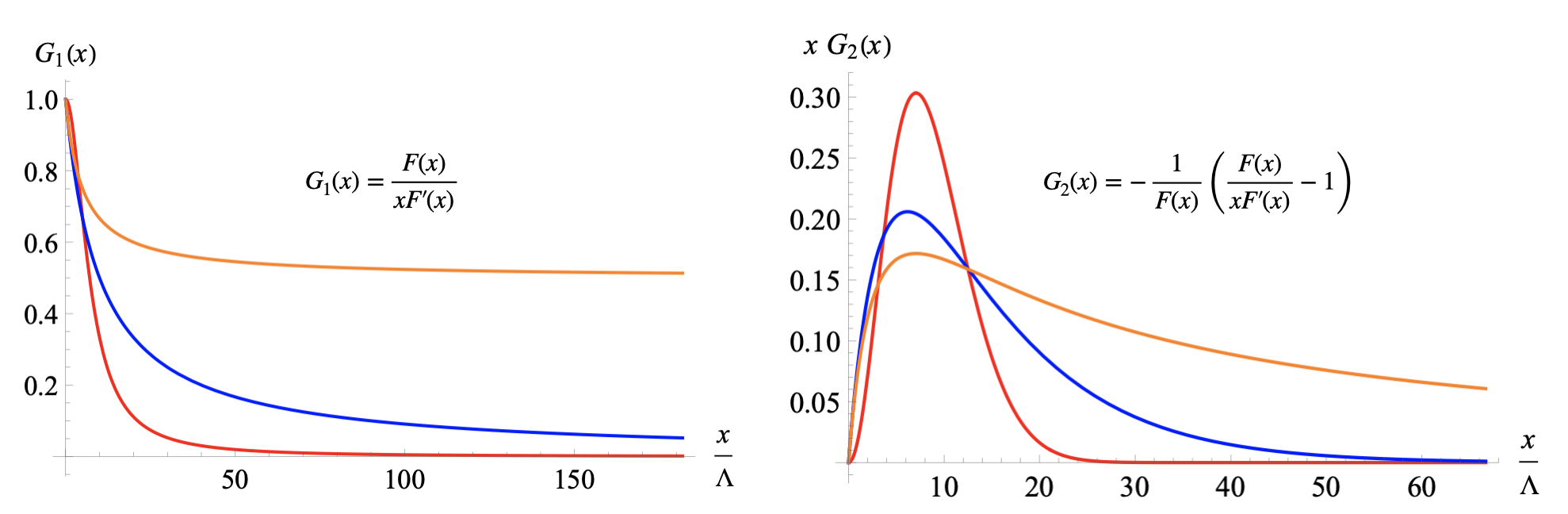} }}%
    \caption {The behaviour of functions $G_1(p^2)$ and $G_2(p^2)$ for three different dispersion relations. These functions are defined by: 
    $\l[{\sf x}^a, {\sf p}_b\r] = i \hbar \l( G_1(p^2) \delta ^a _b+ G_2(p^2) {\sf p}^a {\sf p}_b \r)$. The three curves correspond to $F(x)$:
    (i) $x+x \l[\exp{(x/\Lambda)^2}-1\r]$ (Red), 
    (ii) $x \exp{({x}/{\Lambda})}$ (Blue), 
    (iii) $x\l[1+({x}/{\Lambda})^2\r]$ (Orange).
    Here, $\Lambda$ is a momentum scale and $\Lambda \to \infty$ corresponds to $F(x)=x$. 
}  \label{fig:F3}  %
\end{figure*}

We will now characterise the geometry of the momentum space by starting with the following spacetime picture: two particles of rest energies (masses) $m_1$ and $m_2$ and four momenta $p^i_1$ and $p^i_2$ respectively, intersect at an spacetime event. The energy of this system in it's center-of-momentum frame has the invariant form
\begin{eqnarray}
E_{\rm com}^2 &=& - (p_1+p_2)^2
\nn \\
&=& \mu_1^2 +  \mu_2^2 - 2 \mu_1 \mu_2 {\bm u_1} \cdot {\bm u_2}
\nn \\
&=& \left( \mu_1 +  \mu_2 \right)^2 + 2 \mu_1 \mu_2 \l( \gamma_{\rm rel} -1 \r),
\end{eqnarray}
where we have defined $\bm p_{1,2} = \mu_{1,2} \bm u_{1,2}$ (with ${\bm u}_{1,2}^2=-1$), so that $\bm p^2=-\mu^2$ and $\gamma_{\rm rel}=-\bm u_1 \cdot \bm u_2$ is the relative gamma factor corresponding to the relative velocity $v_{\rm rel}$. 

Note that, conventionally, $\mu$ would be identified with the rest mass $m$ of the particle. However, we postpone this replacement at this stage since we will very soon be interested in generalisation $\mu = f(m)$ which will correspond to the modified dispersion relation $F(p^2)=-m^2$. We will therefore keep the discussion general, but will refer to the case $f(m)=m$ to provide motivation for certain definitions and interpretation.

Consider now the the quantity $\ell^2>0$ defined by:
\begin{eqnarray}
{\ell}^2 &=& 2 \mu_1 \mu_2 \l( \gamma_{\rm rel} -1 \r)
\\
&=& 2 \mu_1 \mu_2 \l( \cosh \chi_1 \cosh \chi_2 - \sinh \chi_1 \sinh \chi_2 \cos \Omega -1 \r),
\nn 
\end{eqnarray}
where the second equality obtains by parameterizing $u^a$ in terms of standard Lorentz transformations: $u^a(\chi, \Omega^A) = \l( \cosh{\chi} \r) T^a + \l( \sinh{\chi} \r) N^a$, where $T^a, N^a$ are arbitrary unit timelike, spacelike vectors in the tangent space ${T}_{{\cal P}_0}(M)$, with $T^a N_a=0$, $\chi$ is rapidity and $\Omega^A=(\theta, \phi)$. Note that, in the non-relativistic limit $v_{\rm rel} \ll c$, and with $\mu_{1,2}=m_{1,2}$, the above expression yields 
$$
\frac{ {\ell}^2 }{2M} \approx \frac{1}{2} \mu_{\rm red} v_{\rm rel}^2,
$$
with $M=m_1+m_2$ and $\mu_{\rm red}=m_1 m_2/M$ (reduced mass). This is therefore the relative kinetic energy of the system. 

We will now take ${\ell}^2$ as the measure of (squared) ``three momentum distance" between the two particles. 
\footnote{Note that one can write $\ell^2=E_{\rm com}^2 - \l[ E_{\rm com}^2 \r]_{v_{\rm rel}=0}$, which has a nice physical interpretation. For a composite system, the 2nd term on the RHS represents the ``rest energy" in the center-of-momentum frame since $v_{\rm rel}=0$. The above relation therefore has the structure ``$|\Vec{p}|^{ 2} = E^2 - m^2$" valid for a point particle.}
Defining $\xi^a = (\mu, \chi, \Omega^A)$ The corresponding metric $g^{3 {\rm {-mom}}}_{ab}$ may be obtained by
\begin{eqnarray}
g^{3-\rm mom}_{ab} &=& \lim \limits_{\xi_2^a \to \xi_1^a} \frac{\partial^2}{\partial \xi_1^a \xi_2^b} \left( \frac{{\ell}^2}{2} \right)
\nn \\
&=& 
\left(
\begin{array}{cccc}
 0 & 0 & 0 \\
 0 & \mu^2 & 0 \\
 0 & 0 & \mu^2 \sinh ^2(\chi ) \sigma_{AB} \\
\end{array}
\right)
\end{eqnarray}
\\
where $\sigma_{AB}$ is the canonical metric on the $2$-sphere. We note that, apart from a zero eigenvalue, the above metric is conformal to the Lobachevsky metric of the relativistic velocity space, $ \DM l_{\rm rel}^2=\DM \chi^2 + \sinh^2 \chi \DM \Omega^2$ \cite{Landau}: $\mu^2 \DM l_{\rm rel}^2=- p^2 \DM l_{\rm rel}^2$. This therefore gives a rigorous justification for our definition of distance measure: it correctly gives a locally Lorentz invariant measure of relative momentum on the space of three momenta. To extend it to a four dimensional metric $\DM \ell^2$, we demand that (i) the four momentum geometry is Lorentzian, and (ii) for points in momentum space that have zero relative velocity, that is, $\DM l_{\rm rel}=0$, the metric gives the difference in rest masses (or rest energies) associated with the corresponding momenta. That is, we require that $\DM \ell^2 = - \DM m^2$ when $\DM l_{\rm rel}^2=0$. This then gives the full metric as:
\begin{eqnarray}
\DM \ell^2 &=& - \DM {m}^2 - p^2 \DM l_{\rm rel}^2
\nn \\
&=& - \frac{p^2 F'^2}{F} \DM \mu^2 + \mu^2 \DM l_{\rm rel}^2,
\end{eqnarray}
where $F'=\DM F/\DM p^2$.

When $F(p^2)=p^2$, we have the standard dispersion relation $p^2=-m^2$ and the above metric is easily recognised as flat Minkowski metric in hyperbolic coordinates, and hence ${\rm \bf Riemann} =0$.

\textit{Modified dispersion relations} which are Lorentz invariant, conventionally given in the form $F(p^2)=-m^2$, correspond to $f(x) = \sqrt{ - F^{-1}(-x^2) }$. These would generically yield ${\rm \bf Riemann} \propto f''/f \neq 0$. In terms of the dispersion relation $F(p^2)=-m^2$, the commutators are given by, 
\begin{eqnarray}
\l[ {\sf x}^{\textsf a}, {\sf p}_{\textsf b} \r] := i \hbar 
\left\{ \delta^{\textsf a}_{\textsf b} + 
\l(\frac{F(p^2)}{p^2 F'(p^2)} - 1 \r) h^{\textsf a}_{\phantom{\textsf a} \textsf b} 
\right\} ,
\end{eqnarray}
where $h^{\textsf a}_{\phantom{\textsf a} \textsf b}$ is the projector orthogonal to $\partial /\partial \mu$. One can write the above expression in a form in which it is easily compared with the existing works on generalised uncertainty principle (for instance, the now familiar version that was introduced in \cite{Kempf, Kempf2}). To do this, one uses the basic definition of normal coordinates, given in Sec. \ref{sec:commutator}, applied to the momentum space geometry, and recognise that $m$ represents the geodesic length (from the origin) in the momentum space metric. The above expression then becomes:
\begin{eqnarray}
\l[ {\sf x}^{\textsf a}, {\sf p}_{\textsf b} \r] &=& i \hbar 
\left( \frac{F}{p^2 F'} \delta^{\textsf a}_{\textsf b} 
- 
\frac{1}{F}\l( \frac{F}{p^2 F'} - 1 \r) {\sf p}^{\textsf a} {\sf p}_{\sf b} 
\right) .
\end{eqnarray}
The remaining commutators then follow from a straightforward application of the Jacobi identity. The final form for the modified Heisenberg algebra is then given by:
\begin{eqnarray}
\l[ {\sf x}^{\textsf a}, {\sf p}_{\textsf b} \r] &=& i \hbar 
\Biggl( G_1(p^2) \delta^{\textsf a}_{\textsf b} + G_2(p^2) {\sf p}^{\textsf a} {\sf p}_{\sf b} \Biggl),
\\
\l[ {\sf p}_{\textsf a}, {\sf p}_{\textsf b} \r] &=& 0,
\\
\l[ {\sf x}^{\textsf a}, {\sf x}^{\textsf b} \r] &=& 2 i \hbar \l( G_2-2G_1'-\frac{2G_1' G_2 p^2}{G_1} \r)\sf x^{[ a} {\sf p}^{b]},
\label{eq:xx-comm}
\end{eqnarray}
where $G_1={F}/(p^2 F')$ and $G_2=-\l(G_1-1\r)/F$. 
(Note, however, that the last commutator above holds for arbitrary 
$G_1, G_2$.) 
Evidently, the modified dispersion relation introduces a non-commutativity in normal coordinates. It will be interesting to study the implications of this non-commutativity vis-a-vis the its connection with modified dispersion \cite{Dispersion relation}. The behaviour of $\l[\sf x^a, p_b \r]$ for some model dispersion relations are shown in the Fig. \ref{fig:F3}. 

\textbf{Some useful limits:} To obtain some useful limits, let us write $F(p^2)=p^2 (1+q(p^2))$ which defines $q(p^2)$. If we demand $q(p^2)$ to be analytic at $p^2=0$, and $q(0)=0$, then $F \sim p^2$ as $p^2 \to 0$ and $F'(0)=1$. In this case, we have 
\begin{eqnarray}
G_1(p^2) &=& 1 - q'(0) p^2 + O(p^4)
\nn \\
G_2(p^2) &=& q'(0) + O(p^2)
\end{eqnarray}
which induces a non-commutativity 
$\l[ {\sf x}^{\textsf a}, {\sf x}^{\textsf b} \r] \sim 6 i \hbar q'(0) \sf x^{[ a} {\sf p}^{b]}$ to the lowest order. 

The opposite limit, $p^2 \to \infty$, depends very much on the fall-off behavior of $G_1, G_2$, which in turn would be determined by the theory that yields the modified dispersion. For instance, an exponential fall-off in, say, $G_2$, of the form $\exp[- \alpha (p^2)^n]$ ($\alpha$ a real positive constant, $n \in$ positive integers) would give a divergent contribution for timelike momenta $p^2<0$ provided $n$ is odd. As a concrete example that does not involve exponentials, consider a dispersion relation of the form $F(p^2)={p^2}({1 + \alpha p^{2 n}})^{-1}$ ($\alpha$ a real constant, $n \in$ positive integers). For such cases, in the high energy limit, $G_1 \sim {1}/{(1-n)}$, while $G_2 \sim \alpha (n/(n-1))  p^{2(n-1)}$. We therefore see that $G_2$ would blow up in the high energy limit for $n>1$. One can do a similar analysis for any dispersion relation; such analysis would be particularly interesting since it might connect analytic properties of $F(p^2)$ with the nature of commutators. 
\section{Discussion} \label{sec:discussion}

We have presented a geometric formalism for the generalised uncertainty principle which is covariant and connects features of the underlying geometry with the deformation of canonical commutator relations. This deformation is tied to extrinsic geometry of the equi-geodesic surfaces of the manifold. When the manifold in question is the momentum space, we characterised its geometry in terms of a four dimensional extension of the relative velocity (Lobachevsky) space, whose Riemann curvature is determined by the modified dispersion relation $F(p^2)=-m^2$. Our work therefore interconnects generalised uncertainty principle, momentum space geometry, and modified dispersion relations in a covariant setting, whose only free function is the one that yields the dispersion relation. 

Let us highlight some key features inherent in the setup we have described: 

As already stated in the Introduction and described in Sec. \ref{sec:commutator}, our choice of variables $({\sf x}^{\textsf a}, {\sf p}_{\textsf b})$ are intrinsically determined by geodesic flows in spacetime and momentum space. These flows are nicely described  in terms of Synge's World function and the van Vleck matrix. Therefore, the construction is completely covariant and in terms of physically well-defined variables.

It is also non-local, since one must choose a base point of the geodesic flow, in both position and momentum space. The non-locality in position space is evident from the corresponding commutators in Eq. (\ref{eq:xx-comm}).

In our description of the momentum space, we have chosen a coordinatization which is based on manifestly covariantly variables, viz the rest mass $m$ and the relative velocity space metric. It will therefore not make much physical sense to talk about general coordinate transformations in the momentum space, although nothing in the formalism forbids it. The equations remain covariant under arbitrary coordinate reparametrizations of the momentum space.

\begin{figure}[!h]%
    {{\includegraphics[width=0.45\textwidth]{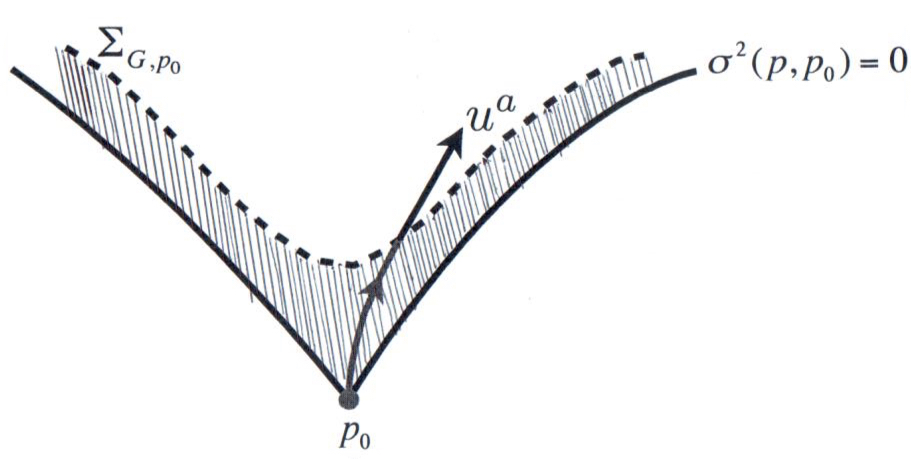} }}%
    \caption {The choice of reference point in the momentum space can be obtained from the vector $u^{\sf a}$ giving normal coordinates in position space.}%
    \label{fig:gen}%
\end{figure}

More interesting is the issue of choice of origin in the momentum space. The astute reader would have noticed that we have taken $m=0$ as the origin. However, we can keep the reference point arbitrary, in which case, the expression for extrinsic curvature $K^{\textsf a}_{\phantom{\textsf a} \textsf b}$ will be messy. Mathematically, one can still compute it, but conceptually, the interpretation is subtle since when choice of origin corresponds to $m \neq 0$, the coordinates in the relative velocity space $(\chi, \theta, \phi)$ must refer to some specific timelike vector. However, note that one does not have to introduce such a vector as an additional object. If ${\sf x} \in I^+({\sf x}_0)$, then the tangent to the geodesic connecting ${\sf x}_0$ to ${\sf x}$ already provides such a vector; see Fig. \ref{fig:gen}. This nicely intertwines the issue of reference point in momentum space with the normal coordinates in the position space, but also makes the former ``position" dependent. It will be worth exploring the physical implications of this. Finally, we wish to highlight that our construction of the momentum space geometry has an elegant interpretation for the standard dispersion relation $p^2=-m^2$: it describes flat spacetime in the Milne coordinates, with rest mass $m$ giving the measure of geodesic length from origin.
%
\section{Acknowledgements} DK acknowledges the support from the Institute of Eminence scheme of IIT Madras funded by the Ministry of Education of India.
        

\appendix*
\section{}
Here, we briefly discuss the commutators between position and momentum operators, their phase-space representations, and constraints arising from the Jacobi identity.  
\begin{equation}
    [\textsf x^{\textsf a}, \textsf p_{\textsf b}]_{\textbf{\textsf p}_0}=i \hbar \Theta^{\textsf a}_{\ \textsf b}(\textbf{\textsf p}_0,\textbf{\textsf p}).
\end{equation}
One can show that the Jacobi identities yields,
\begin{equation}\label{Jacobi identity}
    \left[\textsf p_{\textsf c},[\textsf x^{\textsf a},\textsf x^{\textsf b}]\right]=2i \hbar (\Theta^{[\textsf b}_{\ \textsf c} \textsf x^{\textsf a]}-\textsf x^{[\textsf a}\Theta^{\textsf b]}_{\ \textsf c})=2i\hbar\left[\Theta^{[\textsf b}_{\ \textsf c}, \textsf x^{\textsf a]}\right].
\end{equation}
This clearly indicate that $\l[\sf x^a, \sf x^b \r]$ is not going to commute. Let us try to write the possible representation of the position operator in the momentum basis, then we will give the $[\sf x^a, \sf x^b]$ relation in terms of $\Theta(\textbf{\sf p})$.
\begin{equation}
    \textsf x^{\textsf a}=i\hbar \Theta^{\textsf a}_{\ \textsf m}\partial^{\textsf m}.
\end{equation}
But for being the physical observable the operator must be symmetric with respect to integration measure of the momentum space i.e. $d^4\textsf p\sqrt{-g}$. Using the definition of the symmetric operator,
\begin{equation}
    \int d^4\textsf p\sqrt{- g} \phi^* \hat{\textsf x} \psi=\int d^4\textsf p\sqrt{-g} (\hat{\textsf x} \phi)^*  \psi.
\end{equation}
We write momentum representation of the position operator as,
\begin{equation}\label{position operator}
  \textsf x^{\textsf a}_{\rm sym}=i\hbar(\frac{1}{2}\Theta^{\textsf m \textsf a}\Gamma^{\textsf i}_{\ \textsf m \textsf i}+\frac{1}{2}\Theta_{\textsf m}^{\ \textsf a, \textsf m}+\Theta_{\textsf m}^{\ \textsf a}\partial^{\textsf m}).
\end{equation}
where $f^{,\textsf m}={\partial f}/{\partial \textsf p_{\textsf m}}$. By using the equations (\ref{Jacobi identity}),(\ref{position operator}), we write the  commutation relation between $\hat{\textsf{x}}_{\textsf{i}}$, as 
\begin{equation}
    \left[\textsf x^{\textsf a}, \textsf x^{\textsf b}\right]=i\hbar \{\textsf x^{\textsf l},(\Theta^{-1})^{\textsf m}_{\ \textsf l} \Theta_{\textsf n}^{\ [\textsf a}\Theta_{\textsf m}^{\ \textsf b],\textsf n}\}.
\end{equation}
The \{,\} represent anti-commutators bracket.


\begin{thebibliography}{99}
\bibitem{generators-flows}
This follows simply from the classical picture of a phase space $({\sf x}^{\textsf a}, {\sf p}_{\textsf b})$. A function $\mathscr{G}({\sf x}^{\textsf a}, {\sf p}_{\textsf b})$ on this space defines the vector field 
$$
X_{\mathscr{G}} = \frac{\partial \mathscr{G}}{\partial \sf p_k} \frac{\partial}{\partial \sf x^k} - \frac{\partial \mathscr{G}}{\partial \sf x^k} \frac{\partial}{\partial \sf p_k}
$$
which implies, for any phase space function $\Phi$,
$\delta_{\mathscr G} \Phi = X_{\mathscr{G}}[\Phi] = ( \Phi ,  {\mathscr G})_{\rm PB}$
where ``PB" stands for Poisson brackets. Our identification of the commutators is then the conventional one.
\end{thebibliography}
\end{document}